\documentclass[twocolumn,secnumarabic, amssymb, amsmath, nofootinbib,tightenlines,showpacs, nobibnotes, aps, prb]{revtex4}
\usepackage{amsfonts}

\usepackage{bm}%
\usepackage[colorlinks=true,linkcolor=blue]{hyperref}%
\usepackage{graphicx}
\usepackage{dcolumn}

\begin{document}

\title{Casimir interaction between topological insulators with finite surface band gap}

\author{Liang Chen}
\affiliation{Institute for Theoretical Physics and Department of Modern Physics
University of Science and Technology of China, Hefei, 230026, \textbf{P. R. China}}
\author{Shaolong Wan}
\email[]{slwan@ustc.edu.cn}
\affiliation{Institute for Theoretical Physics and Department of Modern Physics
University of Science and Technology of China, Hefei, 230026, \textbf{P. R. China}}
\date{\today}

\begin{abstract}
Casimir interaction between topological insulators with opposite topological
magnetoelectric polarizabilities and finite surface band gaps has been investigated.
For large surface band gap limit$(m\rightarrow\infty)$, we can obtain results
given in [Phys. Rev. Lett. \textbf{106}, 020403 (2011)]. For small surface band gap
limit$(m\rightarrow0)$, Casimir interaction between topological insulators is
attractive and analogy to ideal mental in short separation limit. Generally, there is a
critical value \(m_c\) and when the surface band gap is greater than the critical
value, the Casimir force is repulsive in an intermediate separation region.
We estimate the critical surface band gap \(m_c \sim 1/(2a)\), where \(a\) is a critical
separation where Casimir force vanishes.
\end{abstract}


\maketitle

\section{Introduction}

\indent Time reversal invariant topological
insulator(TI)\cite{Qi_physToday_2010, Moore_nature_2010,
Hasan_rmp_2010} is a new quantum state of matter which has a full
insulating gap in the bulk, but has gapless surface states
protected topologically. This material has been extensively
studied experimentally\cite{Hsieh_nature_2008,
YLChen_science_2009, YXia_nature_2009, Hsieh_nature_2009,
Hsieh_prl_2009, YLChen_science_2010} and
theoretically\cite{Fu_prl_2007, Fu_prb_2007, Moore_prb_2007,
Qi_prb_2008}. Two dimensional TI has been
observed in \(\text{HgTe}\) quantum
well\cite{Bernevig_science_2006, Konig_science_2007},
\(\text{Sb}_{1-x}\text{Te}_x\) is the first material has been
reported to be 3-dimensional TI, and
\(\text{Bi}_2\text{Se}_3\), \(\text{Bi}_2\text{Te}_3\),
\(\text{Sb}_2\text{Te}_3\) have been
predicted\cite{HJZhang_nature_2009} to be TI
with single Dirac cone on the surface. Novel properties of
TI have been predicted, for instance,
effective monopole\cite{Qi_science_2009} and topological
magnetoelectric effect\cite{Qi_prb_2008}, superconductor proximity
effect induced Majorana fermion states\cite{Fu_prl_2008}
\textit{etc}.

\indent Recently, an interesting property of TI, tunable repulsive Casimir interaction between
TIs with opposite topological magnetoelectric
polarizability \(\theta\) has been proposed\cite{Grushin_PRL}, and
the robustness of this repulsion in small separation limit against
finite temperature and uniaxial anisotropy has also been
analyzed\cite{Grushin_arXiv}. Repulsive Casimir interaction has
been discussed in a few proposals, with special
geometry\cite{Levin_prl_2010} or chiral
metamaterials\cite{Zhao_prl_2009}, or filling high-refractive
liquid between dielectrics\cite{Zwol_pra_2010}. The repulsion
between TIs is analogy to metamaterials, however, time reversal
invariant TI is protected by gapless surface states. In order to
observe the repulsive Casimir interaction, one need cover the TI
surfaces with magnetic coating to open the band gap. The effect of
finite surface band gap on this repulsive force is considerable.

\indent In this paper, we analyze the influence of finite surface band gap on Casimir force between TIs with opposite topological
magnetoelectric polarizability \(\theta\), we show that there is a
minimal surface band gap \(m_c\) and when surface band gap \(m<m_c\),
repulsive Casimir force will disappear. We also estimate this
critical surface band gap numerically.

\indent Let us formulate the model. When time reversal symmetry is
protected in the bulk, the topological nontrivial term
\(\alpha/(4\pi^2)\int{d^3}{x}{d}t\theta\bm{E}\cdot\bm{B}\) can be
reexpressed as spin-momentum locked fermions on the interface of
TI and normal insulator, in this paper we consider only one kind
of fermion corresponding to \(\theta=\pi\) or \(-\pi\),
generalization to multi-fermions is straightforward. Action of
Dirac fermion on TI surface is
\begin{equation}
S_D={\int}d^{3}x\,\bar{\psi}\left[i \gamma^a (\partial_a+ i e A_a) - m \right]\psi,
\end{equation}
where \(a=0,x,y\), \(\gamma^0=\sigma^z\),
\(\gamma^x=iv_{F}\sigma^y\), \(\gamma^y=-iv_{F}\sigma^x\).
\(\sigma^{x,y,z}\) are Pauli matrices of spin, \(v_F\) is the
Fermi velocity of surface fermion, which has a magnitude of
\(10^{-3}\) speed of light(we set \(\hbar=c=1\) in this paper) and
takes different values for different
materials\cite{YLChen_science_2009, YXia_nature_2009},
\textit{ie}, \(v_F=1.3\times10^{-3}\) for
\(\text{Bi}_{2}\text{Te}_{3}\), and \(1.7\times10^{-3}\) for
\(\text{Bi}_{2}\text{Se}_{3}\). Parameter \(m\) is surface band gap
opened by magnetic coating on TI and we assume chemical potential
has been tuned into the surface band gap. \(A_a\) present the first
three components of vector-potential, while electromagnetic field
is described by Maxwell action:
\begin{equation}
\label{eq:S_em_2}
S_{EM} = -\frac{1}{8\pi}{\int}d^4x(\varepsilon\bm{E}^2-\frac{1}{\mu}\bm{B}^2),
\end{equation}
where \(\bm{E}\) and \(\bm{B}\) are electric and magnetic fields,
\(\varepsilon\) and \(\mu\) are permittivity and permeability of TI in the bulk and equal to 1 in the vacuum.

This paper is organized as follows: In Sec. \ref{sec:section2}, we
evaluate an effective action for electromagnetic field on TI
surface by quantum field theory approach and give the Maxwell
equations of electromagnetic field with proper boundary
conditions. In Sec. \ref{sec:section3}, we analyze the Casimir
interaction between TIs via Lifshitz theory. We discuss the
results in Sec. \ref{sec:section4}, and give a conclusion in Sec.
\ref{sec:section5}.

\section{\label{sec:section2}Effective Lagrangian on TI Surface and Maxwell Equations}

In order to calculate the Casimir interaction caused by
quantum fluctuation of electromagnetic field between TIs, one need to
integrate the contribution from surface fermion. An effective
action for external electromagnetic field in (2+1)-dimension can
be found by standard quantum field theory
approach\cite{chay1993,kim1997,novotny2002},
\(S_{eff}(A)=-i\ln\det[i\gamma^a(\partial_a+ieA_a)-m]\). We introduce a Feynman parameter,
integrate out the fermion field up to one-loop correction and get the effective action in the
following form:
\begin{eqnarray}
\label{eq:S_surface_3}
S_{eff}(A) &=& \int{d}^3{x} \left[ - \frac{\phi (\lambda)}{8 \pi} \epsilon_{abc} A^{a} \partial^{b} A^{c} \right. \nonumber\\
&+& \left. \frac{\Phi(\lambda)}{4 {\pi} |m|} \left(F_{0j} F^{0j} +
v_F^{2} F_{xy} F^{xy} \right) \right],
\end{eqnarray}
with dimensionless parameters \(\phi\) and \(\Phi\) which take the
forms:
\begin{eqnarray}
\label{eq:phi}
\phi(\lambda) &=& \text{sign}(m )\alpha \int_0^1{dx} \frac{1}{\sqrt{1 - x ( 1- x) \lambda}}, \\
\label{eq:Phi}
\Phi(\lambda) &=& \alpha \int_0^1{dx} \frac{(1-x) x}{\sqrt{1 - x
(1 - x) \lambda}},
\end{eqnarray}
where \(\text{sign}(m)\) gives the sign of surface band gap, which
corresponding to different signs of topological magnetoelectric
polarizability. \(\alpha=1/137\) is the fine structure constant of
electromagnetic interaction,
\(\lambda=\left[k_0^2-v_F^2\left(k_x^2+k_y^2
\right)\right]/{m^2}\), and \(k_0\), \(k_x\), \(k_y\) are
frequency and momentum of electromagnetic fields on TI surface. A detailed derivation and a short discussion on this effective action(\ref{eq:S_surface_3})  have been given in the appendix. We
also note that in both limit, \(m^2\rightarrow0\) and
\(m^2\rightarrow\infty\), \(\phi\) and \(\Phi\) are convergent.
For the sake of Eq.(\ref{eq:Lifshitz}), we derive expressions of
\(\phi\) and \(\Phi\) in imaginary time formalism:
\begin{eqnarray}
\tilde{\phi} (\gamma) &=& \text{sign}(m) \frac{2 \alpha}{\sqrt{\gamma}} \arctan \left(\frac{\sqrt{\gamma}}{2} \right), \\
\tilde{\Phi} (\gamma) &=& \frac{\alpha}{2 \gamma} + \left(
\frac{\alpha}{4 \sqrt{\gamma}} - \frac{\alpha}{\gamma^{3/2}}
\right) \arctan \left(\frac{\sqrt{\gamma}}{2} \right),
\end{eqnarray}
where \(\gamma=(k_0^2+v_F^2\left(k_x^2+k_y^2\right))/{m^2}\). For
the large surface band gap limit \((|m|\rightarrow\infty)\),
\(\tilde{\phi}(\gamma)\rightarrow{\text{sign}(m)}\alpha\), the
term proportional to \(\phi(\lambda)\) in
Eq.(\ref{eq:S_surface_3}) is topological and the term proportional
to \(\Phi(\lambda)\) in Eq.(\ref{eq:S_surface_3}) is vanishing.
For the small gap limit \((|m|\rightarrow0)\),
\(\tilde{\phi}(\gamma)\rightarrow0\) and
\(\tilde{\Phi}(\gamma)\rightarrow1/6\).

\begin{figure}
\includegraphics[width=5cm]{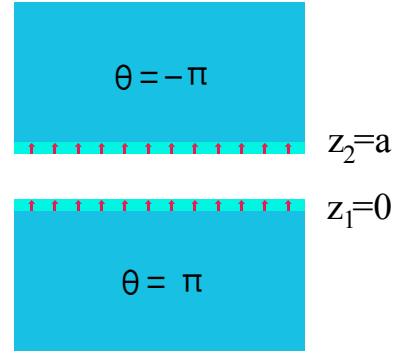}%
\caption{Schematic illustration of Casimir interaction between
TIs with opposite topological magnetoelectric
polarizability \(\theta\). We assume the thickness of magnetic
coating is much smaller than the separation between TIs. \label{fig:setup}}
\end{figure}

\indent Add the surface term Eq.(\ref{eq:S_surface_3}) to standard
action of electromagnetic fields Eq.(\ref{eq:S_em_2}), one can get
the Maxwell equations with surface corrections:
\begin{eqnarray}
&&\frac{1}{4\pi} \nabla \cdot \bm{D} = - \delta (z - z_i) \left(\frac{\phi_i}{4 \pi} B_z - \frac{\Phi}{2 \pi |m|} \nabla \cdot \bm{E} \right), \\
&&\frac{1}{4\pi} \left[\partial_t \bm{D} - (\nabla \times \bm{H}) \right] \nonumber\\
&&= \delta (z  -z_i) \left[\frac{\phi_i}{4 \pi} \tilde{\bm{E}}
+ \frac{\Phi}{2 \pi |m|} \left(\partial_t \bm{E} - v_F^2 \nabla \times \bm{B} \right) \right],\\
&&\nabla \cdot \bm{B} = 0,\\
&&\partial_t \bm{B} + (\nabla \times \bm{E}) = 0,
\end{eqnarray}
where \(\bm{D}=\varepsilon\bm{E}\) and \(\bm{H}=\bm{B}/\mu\) are
electric displacement field and magnetizing field,
\(\tilde{E}_j=\epsilon_{jk}E_{k}\) (\(j,k=x,y\)), (\(i=1,2\)),
\(z_1=0\) and \(z_2=a\) are positions of TI-surfaces (as shown in
Fig.[\ref{fig:setup}]), \(\phi_1\) and \(\phi_2\) are
corresponding values of \(\phi\). Without loss of generality, we
assume the absolute values of surface band gaps on two TIs are equal, different signs of surface band gaps stand for
different signs of the topological term
\(\alpha\theta\bm{E}\cdot\bm{B}/(4\pi^2)\) in Lagrangian of
electromagnetic fields in TIs. We also note
that in large band gap limit \((|m|\rightarrow\infty)\), these Maxwell
equations are equal to those given in Refs.\cite{Qi_science_2009,
Karch_prl_2009} by redefine the electric displacement and
magnetizing field as
\(\bm{D}=\varepsilon\bm{E}+{\alpha}\frac{\theta}{\pi}\bm{B}\),
\(\bm{H}=\frac{1}{\mu}\bm{B}-{\alpha}\frac{\theta}{\pi}\bm{E}\).
From above Maxwell equations, we get the following discontinuous
boundary conditions:
\begin{eqnarray}
D_z (z_i^+) - D_z(z_i^-) &=& - {\phi_i} B_z + \frac{2\Phi}{|m|} \left(\partial_x E_x + \partial_y E_y \right) \\
H_x(z_i^+) - H_x(z_i^-) &=& {\phi_i} E_x - \frac{2 \Phi}{|m|} \left(\partial_t E_y + v_F^2 \partial_x B_z \right)\\
H_y(z_i^+) - H_y(z_i^-) &=& {\phi_i} E_y + \frac{2\Phi}{|m|}
\left(\partial_t E_x - v_F^2 \partial_y B_z \right),
\end{eqnarray}
where \(z_i^{\pm}\) means \(z_i\pm0\). And \(E_x\), \(E_y\),
\(B_z\) are continuous on the interfaces.

\section{\label{sec:section3}Casimir Interaction}

Now we analyze the Fresnel coefficients of reflection light on the
TI-vacuum interface. Incident TE-mode from vacuum with wave-vector
\((k_x, k_y, k_z)\) will induce reflected TE and TM-mode, we
assume the reflection coefficients are \(r_{ee}\) and \(r_{em}\)
respectively, then the electromagnetic waves in the vacuum read:
\begin{eqnarray}
\bm{E} &=& (1 + r_{ee}) k_0 ( -k_y \bm{e}_x + k_x \bm{e}_y ) + r_{em}(-k_z \bm{k} - k^2 \bm{e}_z ), \nonumber\\
\bm{B} &=& (-k_z \bm{k} + k^2 \bm{e}_z) + r_{ee} (k_z \bm{k} + k^2 \bm{e}_z) \nonumber\\
&&+ r_{em} k_0 (- k_y \bm{e}_x + k_x \bm{e}_y ),
\end{eqnarray}
and the refracted light with TE, TM-mode in TI take the forms:
\begin{eqnarray}
\bm{E}&=&t_{ee} k_0(-k_y\bm{e}_x + k_x\bm{e}_y )+c\,t_{em}(p_z\bm{k}-k^2\bm{e}_z),\nonumber\\
\bm{B}&=&t_{ee}(-p_z\bm{k}+k^2\bm{e}_z) + \frac{t_{em}}{c}k_0( -k_y \bm{e}_x + k_x \bm{e}_y ),
\end{eqnarray}
where \(t_{ee}\) and \(t_{em}\) are refraction coefficients of TE
and TM-mode, \(c\) is the relative velocity of light in TI bulk,
\(\bm{k}=k_x\bm{e}_x+k_y\bm{e}_y\), \(k^2=k_x^2+k_y^2\) and
\(p_z\) is \(z\)-component of wave vector in TI. For the injected
TM-mode, one can write the analogy equations with reflection
coefficients \(r_{me}\), \(r_{mm}\) and refraction coefficients
\(t_{me}\), \(t_{mm}\). After some tedious derivation, we obtain
the Fresnel coefficients matrix \(\mathcal{R}\) in imaginary time
formalism:
\begin{equation}
\label{eq:fresnel_coefficients}
\mathcal{R}=\left(
  \begin{array}{cc}
    r_{ee} & r_{em} \\
    r_{me} & r_{mm} \\
  \end{array}
\right),
\end{equation}
with
\begin{eqnarray}
r_{ee}&=&-1+\frac{2}{D}\left(1+\varepsilon\frac{k_z}{p_z}+2\tilde{\Phi}\frac{k_z}{m}\right),
\nonumber\\
r_{em}&=&r_{me}=\frac{2}{D} \tilde{\phi}, \nonumber\\
r_{mm}&=&1-\frac{2}{D}\left(1+\frac{1}{\mu}
\frac{p_z}{k_z}+2\lambda\tilde{\Phi}\frac{m}{k_z}\right),
\nonumber\\
\end{eqnarray}
where the denominator
\begin{eqnarray}
D& & = \left(1+\varepsilon\frac{k_z}{p_z} \right)
\left(1+2\gamma\tilde{\Phi} \frac{m}{k_z}\right) +
\left(1+\frac{1}{\mu }\frac{p_z}{k_z}\right) \nonumber\\
& &\times\left(1+2\tilde{\Phi} \frac{k_z}{m}\right)
+\left(\frac{\epsilon }{\mu }+\tilde{\phi} ^2\right)-\left(1-4
\gamma\tilde{\Phi} ^2\right).
\end{eqnarray}

For the large surface band gap limit, we can obtain the same Kerr
rotation and Faraday rotation angle as given in
Ref.\cite{Qi_prb_2008, Karch_prl_2009}.

In imaginary time formalism, Casimir energy density between two
parallel dielectric semispaces can be expressed in a closed form
of dielectric permittivity:
\begin{equation}
\label{eq:Lifshitz} \frac{E_C(a)}{A}=\int_0^{\infty} \frac{d
k_0}{2\pi} \int \frac{{d}^2k_{\parallel}}{(2 \pi)^2}
\log\det\left[ 1 - \mathcal{R}^{(1)} \mathcal{R}^{(2)} \bm{e}^{-2
k_3 a} \right]
\end{equation}
where \(A\) is the surface area of TIs, \(\mathcal{R}^{(1,2)}\)
are Fresnel coefficients on the surfaces,
\(k_3=\sqrt{\bm{k}_{\parallel}^2+k_0^2}\). In order to calculate
the Casimir energy density numerically, we also need a form of
frequency-dependent dielectric permittivity \(\varepsilon\) (we
assume the permeability \(\mu = 1\)), this can be modeled
by\cite{Bordag_book, Bordag_report_2001}:
\begin{equation}
\label{eq:dielectric_permittivity}
\varepsilon(i k_0) = 1+\sum_{J=1}^{K}\frac{g_J}{k_{0}^{2}+\omega_{J}^{2}+\gamma_{J} k_0},
\end{equation}
we consider only one oscillator (\(K=1\)) with oscillator
strength \(g_J\),  oscillator frequency \(\omega_{J}\) and
damping parameter \(\gamma_{J}\). \(\gamma_{J}\ll\omega_J\) and we
omit the contribution from damping parameter here.

\section{\label{sec:section4}Results and Discussion}

\begin{figure}
\includegraphics[width=9cm]{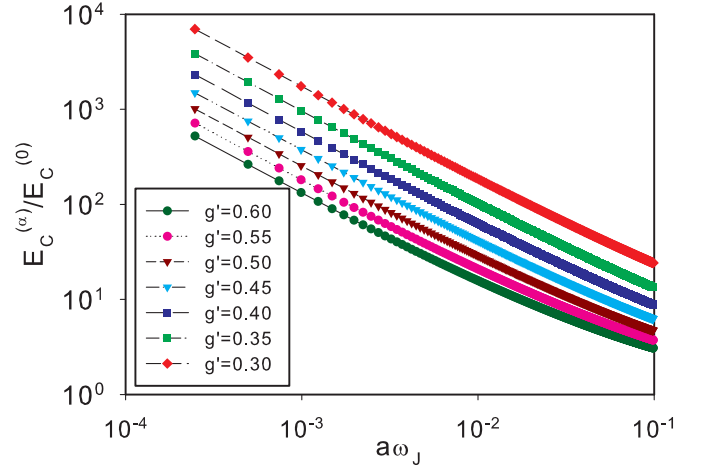}%
\caption{The ratio \(E_C^{(\alpha)}/E_C^{(0)}\) as a function of
dimensionless separation \(a \omega_J\) for different oscillator
strength \(g' = \sqrt{g_J}/\omega_J\) in the closed surface band gap
limit, \(m = 0\). Where \(E_C^{(\alpha)}\)(\(E_C^{(0)}\)) is the
Casimir energy with(without) surface correction. Here fermion
velocity \(v_F = 1.0\times10^{-3}\).\label{fig:zero_mass}}
\end{figure}

In this paper, we analyze Casimir interaction between TIs
with finite surface band gap. First, for large surface band gap
limit \((m \rightarrow \infty)\), we can obtain same results
given in [Phys. Rev. Lett. \textbf{106}, 020403 (2011)]\cite{Grushin_PRL} from equations
(\ref{eq:fresnel_coefficients})-(\ref{eq:dielectric_permittivity}).
Second, for small surface band gap limit \((m \rightarrow 0) \), the
off-diagonal terms in Fresnel coefficients matrices will vanish
and Casimir energy can be rewritten in imaginary time formalism
as:
\begin{eqnarray}
\label{eq:Lifshitz_zero_mass} \frac{E_C(a)}{A} &=& \int_0^\infty
\frac{{d}k_0}{2 \pi} \int \frac{d^2k_{\parallel}}{(2 \pi)^2}
\left[\log \left(1 - e^{-2 k a} r_{TE}^{(1)}r_{TE}^{(2)} \right)
\right.
\nonumber\\
&&+ \left.\log \left(1 - e^{-2 k a} r_{TM}^{(1)} r_{TM}^{(2)}
\right) \right],
\end{eqnarray}
with
\begin{eqnarray}
\label{eq:re_zero_mass}
r_{TE} &=& - 1 + \frac{2}{1 + \frac{p_3}{k_3} + \frac{\pi \alpha}{4} \sqrt{\cos^2 \theta + v_F^2 \sin^2 \theta}},\\
\label{eq:rm_zero_mass} r_{TM} &=& 1 - \frac{2\frac{p_3}{k_3}
\sqrt{\cos^2 \theta + v_F^2 \sin^2 \theta}}{\frac{\pi \alpha}{4}
\frac{p_3}{k_3} + (\frac{p_3}{k_3} + \varepsilon) \sqrt{\cos^2
\theta + v_F^2 \sin^2 \theta}} ,
\end{eqnarray}
where \(k_3=\sqrt{k_0^2+\bm{k}_{\parallel}^2}\),
\(p_3=\sqrt{\varepsilon{k_0^2}+\bm{k}_{\parallel}^2}\), and
\(\theta=\cos^{-1}(k_0/k_3)\).

\indent The Casimir energy between dielectric materials without
special boundary conditions, \(\alpha\rightarrow0\) in
Eq.(\ref{eq:re_zero_mass}) and Eq.(\ref{eq:rm_zero_mass}), has
been studied\cite{Bordag_book, Bordag_report_2001, Lambrecht_NJP}.

\indent Considering correction from surface interaction, for large
separation limit, we obtain the correction up to first order of
fine structure constant:
\begin{eqnarray}
&&\frac{E_C^{(1)}(a)}{E_0} = - \frac{\pi \alpha}{4 d^3}
\left[\frac{\varepsilon(0) - 1}{(\varepsilon(0)
+ 1)^3} \log{\frac{1}{v_F}} \right. \nonumber\\
&&+ \left. \frac{\log \left[\frac{1}{2} \left(1 +
\sqrt{\varepsilon(0)} \right) \right]}{\varepsilon(0) - 1} -
\frac{3 + 5 \sqrt{\varepsilon(0)}}{4 \left(1 +
\sqrt{\varepsilon(0)} \right)^3} \right],
\end{eqnarray}
where \(E_0=A\omega_J^3/(2\pi)^2\) which is set as the unit of
Casimir energy, \(d=a\omega_J\) is the dimensionless separation.

\indent For small separation limit, in order to make the physics
more clear, we also formally expand
Eq.(\ref{eq:Lifshitz_zero_mass}) in powers of \(\alpha\), up to
first order correction, the Casimir energy takes the following
form(here we assume the relative oscillator strength
\(g_J/\omega_J^2\ll1\)):

\begin{eqnarray}
\frac{E_C^{(1)}(a)}{E_0} & &= - \frac{g_J}{\omega_J^2} \frac{\pi
\alpha}{64 d^3} \int_0^{\infty} d
y y^2 e^{- y} \nonumber\\
&&\left[\frac{\theta(t)}{\sqrt{t}} \text{arctan} \sqrt{t} +
\frac{\theta (-t)}{\sqrt{-t}} \text{arctanh} \sqrt{-t} \right],
\end{eqnarray}
where \(t=-1+{v_F^2y^2}/{4d^2}\) and \(\theta(t)\) is the
Heaviside unit step function. Casimir energy is dominated by
surface Dirac fermion and turns into the ideal conductor case
which is proportional to \(1/a^3\). This conclusion is also
confirmed numerically in Fig.[\ref{fig:zero_mass}].
\begin{figure}
\includegraphics[width=9cm]{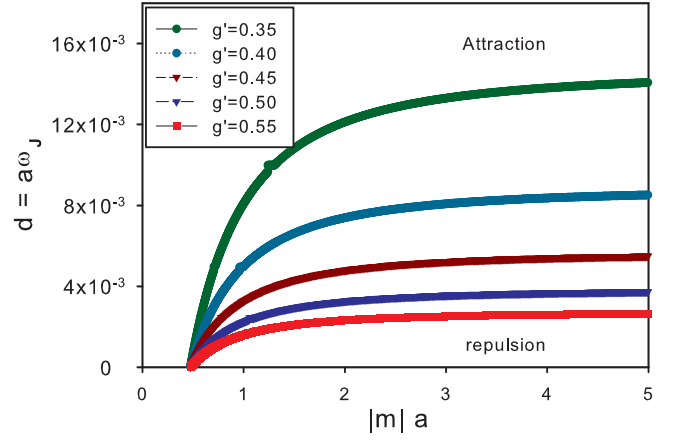}%
\caption{Boundary of repulsive and attractive Casimir interaction in the plane of
dimensionless separation \(d=a\omega_J\) and product \(|m|a\) for
different oscillator strengths \(g'=\sqrt{g_J}/\omega_J\). When
the parameters \(d\) and \(|m|a\) have been taken in the left-up
region over these lines, the Casimir interaction is attractive,
when \(d\) and \(|m|a\) have been taken in the lower-right region
of these lines, the Casimir interaction is repulsive. (The
relative Fermi velocity \(v_F\) has been taken to be
\(1.0\times10^{-3}\)).\label{fig:phase_boundary}}
\end{figure}

\indent Finally, for general surface band gap, we have two
dimensionless parameters: \(m/\omega_J\) and \(d=\omega_J{a}\)
(there are two other parameters in our model, the Fermi velocity
of surface fermion, \(v_F\), and optical oscillator strength in
TIs, \(g_J/\omega_J^2\), which both have quantitatively influence
on Casimir energy). For the large separation limit
\((a\gg\text{max}(1/\omega_J,1/|m|))\), we expand the integral in
Eq.(\ref{eq:Lifshitz}) in power of fine structure
constant\cite{alpha} \(\alpha\) and
consider the correction up to \(\alpha\). In this case, the
dielectric permittivity \(\varepsilon(ik_0)\) can be approximated
by long wave length limit value \(\varepsilon(0)\), and the
Casimir energy correction from interaction between surface
fermions and electromagnetic field reads:

\begin{equation}
\frac{E_{C}^{(1)}(a)}{E_0} = - \frac{|m| \alpha}{\omega_{J}d^2}
\int_0^1 \frac{dx}{v_F^2+x^2} \left[\frac{r^2(\varepsilon(0) -
r)}{(\varepsilon(0) + r)^3} + \frac{x^2(1 - r)}{(1 + r)^3} \right],
\end{equation}
where \( r= \sqrt{1 + (\varepsilon(0)-1) x^2}\), and \(v_F \ll
1\).

\indent For the small separation limit \((d\rightarrow0)\), in
order to make the physics more clear, we also formally expand the
Casimir energy in power series of \(\alpha\). In this case, the
Casimir energy is dominated by surface terms, the term which contains
\(\tilde{\Phi}^2\) and is proportional to \(1/m^2a^5\) is
important. However this dominant term will be suppressed if
\(|m|a\rightarrow\infty\), the topological term which contains
\(\text{sign}(\theta_1\theta_2)\tilde{\phi}^2\) and is proportional
to \(1/a^3\) will provide a large repulsive potential between TIs
when \(\text{sign}(\theta_1\theta_2)=-1\). So the surface terms in
Casimir energy will dominate and \(|m|a\) is a good parameter to
estimate the Casimir force: when \(|m|a\gg1\), the Casimir force
will be repulsive and when \(|m|a\ll1\), the Casimir force will be
attractive.

\begin{figure}
\includegraphics[width=8.5cm]{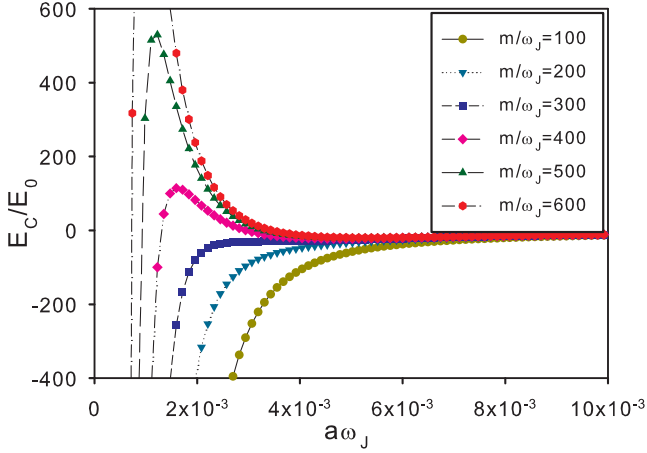}%
\caption{Casimir energy density \(E_C\) (in units of
\(E_0=\omega_J^2/(2\pi)^2\)) as a function of the dimensionless
distance \(d=a\omega_J\) with different surface band gaps
\(m/\omega_J\). Here we take the dimensionless oscillator strength
\(g_J/\omega_J^2=0.45^2\) and Fermi velocity
\(v_F=1.0\times10^{-3}\).\label{fig:finite_gap}}
\end{figure}

\indent In Fig.[\ref{fig:phase_boundary}], we give the boundary
of repulsive and attractive Casimir interaction, as a
function of dimensionless separation \(d=a\omega_J\) and product
\(|m|a\). We find that there is a critical value
\((|m|a)_{c}\sim1/2\), when \(|m|a<(|m|a)_{c}\), the Casimir
interaction is attractive for any separation length. The
independence of \((|m|a)_{c}\) on oscillator strength
\(g_J/\omega_J^2\) shows that Casimir interaction, in small
separation limit, is dominated by surface terms. More intuitively,
we calculated the Casimir energy as a function of dimensionless
separation \(d=a\omega_J\) for different surface band gaps, as shown in
Fig.[\ref{fig:finite_gap}], for given parameters,
\(g_J/\omega_J^2=0.45^2\) and \(v_F=1.0\times10^{-3}\). We find the critical surface band gap, where the repulsive peak disappears, \(m_c\approx300\omega_J\)
(the blue-square dotted line in Fig.[\ref{fig:finite_gap}]).

\begin{figure}
\includegraphics[width=9cm]{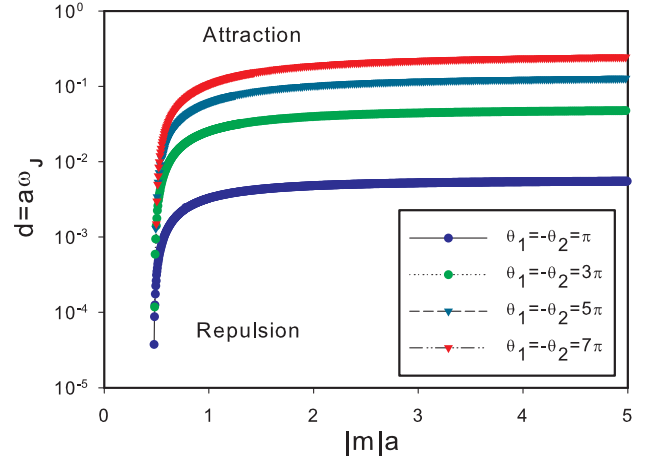}%
\caption{Boundary of repulsive and attractive Casimir interaction in the plane of
dimensionless separation \(d=a\omega_J\) and product \(|m|a\) for
different topological magnetoelectric polarizability. (The
relative Fermi velocity \(v_F\) and oscillator strength
\(g_J/\omega_J^2\) has been taken to be \(1.0\times10^{-3}\) and \(0.45^2\)
respectively).\label{fig:phase_boundary_pi}}
\end{figure}

\indent We note that our calculations can be generalized to multi-value of
topological magnetoelectric polarizability \(\theta=(2n+1)\pi\) ($n$
is an integer) straightforward by introducing multi-fermion on TI
surface, and the critical value \((|m|a)_{c}\) is independent on the
absolute value of $\theta$ (as shown in Fig.[\ref{fig:phase_boundary_pi}]),
this is because in short separation limit, Casimir interaction is dominated
by surface terms and each species fermion will contribute both repulsive and attractive Casimir interaction if $\text{sign}(\theta_1) = - \text{sign}(\theta_2)$.

\indent We can use this relationship to estimate the critical
surface band gap for repulsive Casimir interaction. For \(\text{Tl}
\text{Bi} \text{Se}_2\) suggested in Ref.\cite{Grushin_PRL}, the
minimum of Casimir energy appears at a separation of \(a \sim
0.1{\mu}m\), and the corresponding surface band gap needs to be greater
than \(1eV\), which reflects that the width of surface band gap opened
by magnetic coating is non-ignorable and unaccessible
experimentally.

\section{\label{sec:section5}Conclusion}

We studied the Casimir energy between TIs with
opposite topological magnetoelectric polarizability and finite
surface band gap via Lifshitz formula, we found that, in small
separation limit, Casimir force is dominated by interaction
between surface fermion and electromagnetic field in the vacuum,
and a great surface band gap \(m>m_c\sim1/(2a)\) is essential for
repulsive Casimir interaction.

\begin{acknowledgments}
This work is supported by NSFC Grant No.10675108.
\end{acknowledgments}

\appendix*
\section{Effective action}
\label{app:effective_action}
We give a detailed derivation of the effective action(\ref{eq:S_surface_3}) in this appendix.
The effective action from quantum field theory is:
\begin{equation}
\label{app:S_eff}
S_{eff}(A) = \frac{1}{2}\int \frac{d^3k}{(2\pi)^3} A_{a}(k)\Pi^{ab}(k)A_{b}(k),
\end{equation}
where \(\Pi(k)\) is the polarization tensor, which takes the form:
\begin{equation}
\label{app:polar}
i\Pi^{ab}(k) = -e^2\int\frac{d^3p}{(2\pi)^3}\text{tr}[(-i\gamma^a)G(k+p)(-i\gamma^b)G(k)],
\end{equation}
and \(G(k) = i/(\gamma^a k_a + m)\) is the propagator of fermion on TI surface.
From the standard calculation in quantum field theory, one can get the exact form of polarization tensor:
\begin{eqnarray}
\Pi(k) &=& \Pi_1(k) + \Pi_2(k)\\
\label{app:Pi_1}
\Pi_1^{ab}(k) &=& \frac{\phi(\lambda)}{4\pi}\epsilon^{abc}ik_c\\
\label{app:Pi_2}
\Pi_2(k) &=& \frac{\Phi(\lambda)}{2\pi|m|}\left(
\begin{array}{ccc}
 k_x^2+k_y^2 &  -k_0 k_x &  -k_0 k_y \\
  -k_0 k_x & k_0^2- v_F^2 k_y^2 & v_F^2 k_x k_y  \\
  -k_0 k_y & v_F^2 k_x k_y  & k_0^2-v_F^2 k_x^2
\end{array}
\right)\nonumber\\
\end{eqnarray}
where \(\phi(\lambda)\) and \(\Phi(\lambda)\) has been given in Eq.(\ref{eq:phi}) and Eq.(\ref{eq:Phi}), \(k_{1,2}(k_0)\) are the momentum(frequency)
of electromagnetic field. One can check that the polarization tensor satisfies Ward identity, \(\sum_a k_a \Pi^{ab}(k) = \sum_b\Pi^{ab}(k) k_b = 0 \).
The Fourier transformation of Eq.(\ref{app:S_eff}) gives Eq.(\ref{eq:S_surface_3}).\\
\indent We take \(\Pi^{xy}(k)\) as an example to show more detailed calculations of polarization tensor. Taking the trace in Eq.(\ref{app:polar}), one can get
\begin{widetext}
\begin{equation}
i\Pi^{xy}(k)
=-e^2 \int \frac{d^3p}{(2\pi)^3}\frac{2v_F^2[-im k_0 + v_F^2(2k_x k_y + k_x p_y + k_y p_x)]}
{[(p_0 + k_0)^2 + m^2 -v_F^2(\bm{p}+\bm{k})^2][k_0^2 + m^2 -v_F^2\bm{k}^2]}.
\end{equation}
\end{widetext}
One can get the following form of \(i\Pi^{xy}(k)\) by introducing a Feynman parameter \(x\) and redefining the integration variables \(l_a '=p_a+xk_a\), \(l_0 = l_0 '\), and \(\bm{l}=v_F\bm{l}'\):
\begin{equation}
i\Pi^{xy}(k) =-2e^2\int_0^1dx\int\frac{d^3l}{(2\pi)^3}\frac{i m k_0+2x(1-x)v_F^2k_xk_y}{\left(l_0^2-\bm{l}^2-\Delta\right)^2},
\end{equation}
where \(\Delta= m^2-x(1-x)(k_0^2-v_F^2\bm{k}^2)= m^2[1-\lambda x (1-x)]\).
Making Wick rotation \(l_0\rightarrow{il_0^E}\) and integration over \(l\), we find:
\begin{eqnarray}
i\Pi^{xy}(k)&=&ie^2\left[imk_0\int\frac{ d x }{4\pi\sqrt{\Delta}}+v_F^2k_xk_y\int dx\frac{x(1-x)}{2\pi\sqrt{\Delta}}\right]\nonumber\\
&=&i\left(\frac{i k_0}{4\pi}\phi(\lambda)+\frac{v_F^2k_xk_y}{2\pi|m|}\Phi(\lambda)\right)
\end{eqnarray}
\indent Comparing with the effective action of electromagnetic field in monolayer graphene system as shown in Ref.\cite{graphene},
we find that there is an additional topological term Eq.(\ref{app:Pi_1}) together with the normal vacuum polarization Eq.(\ref{app:Pi_2}),
the first term is essential for TI because this parity-odd term reflects the fact that there are always odd species of surface fermions
which are spin-momentum locked, the contribution from second term is analogy to Dirac fermion in monolayer graphene system and reflects
the dynamical response of TI surface state to extra electromagnetic field.\\


\begin{thebibliography}{arxiv}
\bibitem[31]{alpha}
                We note that we do not expand the integral in power of $1/d$, the reason is, in short separation limit
                integration over momentum will give a divergence which set the term proportional to $\alpha^2$ large than next
                to leading order term or even the leading order term (in large surface band gap limit), this divergence is supressed
                in large separation limit.
   \label{alpha}
\end{thebibliography}

\end{document}